\documentclass{statsoc}

\usepackage[a4paper]{geometry}
\usepackage{graphicx}
\usepackage[textwidth=8em,textsize=small]{todonotes}
\usepackage{amsmath}
\usepackage{natbib}
\usepackage[caption=false]{subfig}

\title[Sequential designs using Bernstein kernels]{Model selection for sequential designs in discrete finite systems using Bernstein kernels}
\author[M. Nath and S. Eubank]{Madhurima Nath $^{\footnote{{\it{Address for correspondence}}: NDSSL, Biocomplexity Institute of Virginia Tech, 1015 Life Science Circle, Blacksburg, VA 24061, USA}}$ $^{1, 2}$ and Stephen Eubank$^{1,2,3}$}
\address{$^{1}$Network Dynamics and Simulation Science Laboratory, Biocomplexity Institute of Virginia Tech, $^{2}$Department of Physics, $^{3}$Department of Population Health Sciences, Virginia Tech, Blacksburg, Virginia 24061, USA.}
\email{\{mnath,eubank\}@vt.edu}

\begin{document}
\begin{abstract}
We view sequential design as a model selection problem to determine which new observation is expected to be the most informative, given the existing set of observations. 
For estimating a probability distribution on a bounded interval, we use bounds constructed from kernel density estimators along with the estimated density itself to estimate the information gain expected from each observation.
We choose Bernstein polynomials for the kernel functions because they provide a complete set of basis functions for polynomials of finite degree and thus have useful convergence properties.
We illustrate the method with applications to estimating network reliability polynomials, which give the probability of certain sets of configurations in finite, discrete stochastic systems.
\end{abstract}


\section{Introduction}
The probability of conf\/igurations in a f\/inite, discrete, stochastic dynamical system can be represented as a histogram with a f\/inite number,  $N+1$, of bars.
The $N+1$ bar heights are completely determined by a degree $N$ polynomial, but determining the coef\/f\/icients of the monomial terms in the polynomial from data is notoriously ill-conditioned.
Here we develop a numerically stable approach using an alternative basis of Bernstein polynomials instead of monomials and apply it to designing an ef\/f\/icient experiment to estimate the probability of conf\/igurations.

Given
\begin{enumerate}
\item a function $f(x)$ that can be represented as an $N$ degree polynomial and
\item an ``oracle" that can evaluate the function for arbitrary $x$, but at a certain cost $C$ for each query, 
\end{enumerate}
we would like to f\/ind
\begin{enumerate}
\item the sequence of points $x_m$ for querying the oracle such that each point gives the maximum amount of information about the function, i.e., most reduces the estimation error and
\item a stopping criterion that lets the user decide when the amount of information to be gained is not worth the cost of calling the oracle, $C$ again.
\end{enumerate}

We can project any degree $N$ polynomial $f(x)$ uniquely onto the Bernstein basis of order $N$, 
\begin{equation}
f(x) = \sum_{k=0}^N \beta_k B(N,k,x),
\label{eqn:bern}
\end{equation}
where $B(N,k,x)=\binom{N}{k} x^k (1-x)^{N-k}$ is the $k$th Bernstein basis function of order $N$ and $\beta_k$ are the Bernstein or B\'{e}zier coef\/f\/icients \citep{bernvstein1912demonstration, levasseur1984probabilistic, lorentz2012bernstein}. 
The function $f(x)$ can thus be represented as a B\'{e}zier curve with control points $(k,\beta_k)$.
B\'{e}zier curves are frequently used in computer graphics, \citep{bezier1974mathematical, farin1983algorithms, bourke1996bezier} in part because of the ef\/f\/iciency of de Casteljau's recursive algorithm. It has been shown that de Casteljau's algorithm is a numerically stable method for evaluating B\'{e}zier curves at arbitrary parameter values \citep{de1986shape, chang1994generalized, phillips1997casteljau}. 
A B\'{e}zier curve is completely contained in the convex hull of its control points and it always begins at the f\/irst control point and ends at the last one. 
%Certain properties of the Bernstein polynomials like:
%\begin{itemize}
%\item $B(N,k,1-f\/ix) = B(N, N-k, x)$
%\item $\int_0^1 B(N,k,x) dx = \frac{1}{N+1}$, $\forall k = 1, \ldots , N$
%\end{itemize}
%prove to be extremely useful to select the optimal set of points for estimating $f(x)$. 

Bernstein polynomials have been extensively used to approximate a bounded and continuous function and adapted for smooth estimation of a distribution function concentrated on the interval $[0,1]$ \citep{JOGESHBABU2002377, comte2011data, daouia2016data}. It has been shown that Bernstein estimators converge to the true densities for a degree $N$ polynomial \citep{petrone2002consistency}.

The function $f$ depends on the dynamical system. Further, the optimal $(m+1)$-th measurement will typically depend on both the previous points and the values of the function at those points, making this an adaptive method. When the function $f$ is unknown, there is no {\em guarantee} of obtaining the most information possible out of every choice, but we propose a process that bounds the expected value of the information that is gained, given what we have already learned. Stopping criteria are determined by the Akaike Information Criterion ($AIC$) or Bayesian Information Criterion ($BIC$) applied to the  \emph{L2 norm} of successive estimates for $f$. \emph{L2 norm} provides our best estimate of the error in $\widehat{f}$. If the error from the f\/it is within acceptable limits, we stop calling the oracle for additional points. The values of $AIC$ and $BIC$ give the information content of the estimator. Given a set of competing estimators to f\/it $f(x)$, the one with minimum $AIC$ and $BIC$ is preferred \citep{akaike1974new, akaike1998information, schwarz1978estimating}. We can view the model generated by each possible choice of measurement as a competing estimator and choose the measurement that provides the most marginal information. We will demonstrate these processes with two examples for which the function $f$ is known and consider the effect of estimation error in the $\beta_k$ coef\/f\/icients themselves.

\section{Method}
For ease of explication, we focus on estimating a cumulative probability function of a single variable $x\in[0,1]$. Hence $f$ is monotonic non-decreasing, with $f(0)=0$ and $f(1)=1$.
Motivated by the typical behavior of the reliability polynomial for many stochastic systems, and without loss of generality, 
we introduce two parameters, $k_{min}$ and $k_{max}$ in our analysis, def\/ined as follows: $k_{min}$ is the minimum $k$ for which $\beta_k > 0$ and $k_{max}$ is the maximum $k$ such that $\beta_k < 1$. 
The corresponding $\beta$s are $\beta_{k_{min}}$ and $\beta_{k_{max}}$. 
In many applications, there are ef\/f\/icient methods for determining $k_{min}$, $k_{max}$ and often their associated $\beta$s, e.g., shortest path algorithms, minimum cut algorithms, and the Kirchof\/f matrix tree theorem.
Out of the $N$ coef\/f\/icients, now $n = N-k_{min}-(k_{max} + 1)$ are unknown while the rest are $0$s, $1$s, $\beta_{k_{min}}$ and $\beta_{k_{max}}$.

We call the estimator obtained with just these 4 parameters -- $(k_{min}, \beta_{k_{min}})$ and $(k_{max}, \beta_{k_{max}})$ -- the ``estimator with no other knowledge". There is an $n$-parameter family of models consistent with these parameters. Bounds on $f(x)$ can be obtained by choosing extremal values for the unknown $\beta_k$s. A reasonable estimator, $\widehat{f}(x)$, can be obtained by linearly interpolating the unknown coef\/f\/icients between the points $(k_{min}, \beta_{k_{min}})$ and $(k_{max}, \beta_{k_{max}})$.
The estimator $\widehat{f}(x)$ is often surprisingly good.
To improve the f\/it, the oracle is called to give a measurement of one of the unknown $\beta_k$s and the remaining coef\/f\/icients are again estimated using piecewise linear interpolation between the known coef\/f\/icients.
Knowledge of one more $\beta_k$ improves both the goodness-of-f\/it for $\widehat{f}(x)$ and the bounds on $\widehat{f}(x)$. 
The optimization process allows us to select which measurement is likely to reduce the f\/itting error most.
The process can be repeated till either all $\beta_k$s have been measured by the oracle or the stopping criterion is reached.

We label the different estimators used in the process of estimating the exact $f$ as $\widehat{f}_{m}$, where $m = 0, 1,2, \ldots, n$ gives the number of additional known $(k, \beta_k)$. $\widehat{f}_{0}$ is the ``estimator with no additional knowledge", $\widehat{f}_{1}$ is $\widehat{f}_{0}$ with one additional coef\/f\/icient known i.e., $(3, \beta_3)$ in the toy example and so on. The goodness-of-f\/it is calculated by evaluating the \emph{L2 norm} of $f(x)-\widehat{f}(x)$:
\begin{align}
\|f(x)-\widehat{f}(x)\|_2 = & \sqrt{\int_0^1 \left[ f(x) - \widehat{f}(x) \right]^2 dx} \\ 
= & \sqrt{ \frac{1}{2N+1} \left( \sum_{i=0}^N \sum_{j=0}^N \left( \beta_i - \widehat{\beta}_i \right) \left(\beta_j - \widehat{\beta}_j \right) \frac{\binom{N}{i} \binom{N}{j}}{\binom{2N}{i+j}} \right) } \label{eqn:norm}
\end{align}
where $\widehat{\beta}$s are the linearly interpolated $\beta$'s used to calculate $\widehat{f}(x)$. The integral of the Bernstein basis functions gives the factor $\frac{1}{2N+1}$. It is to be noted that $\|f(x)-\widehat{f}(x)\|_2$ is not simply the root-mean-squared error in the $\beta_k$ coef\/f\/icients themselves.
Cross terms appear because the Bernstein polynomials are not orthogonal. Also, the residuals for $f(x)$ are the weighted smoothened version of those for $\beta_k$.
If the exact $f(x)$ is unknown, we can nonetheless obtain provable bounds with the knowledge of the initial constraints and the values of $(k_{min}, \beta_{k_{min}})$ and $(k_{max}, \beta_{k_{max}})$. 

\begin{figure}[ht!]
\centering
\includegraphics[width=\linewidth]{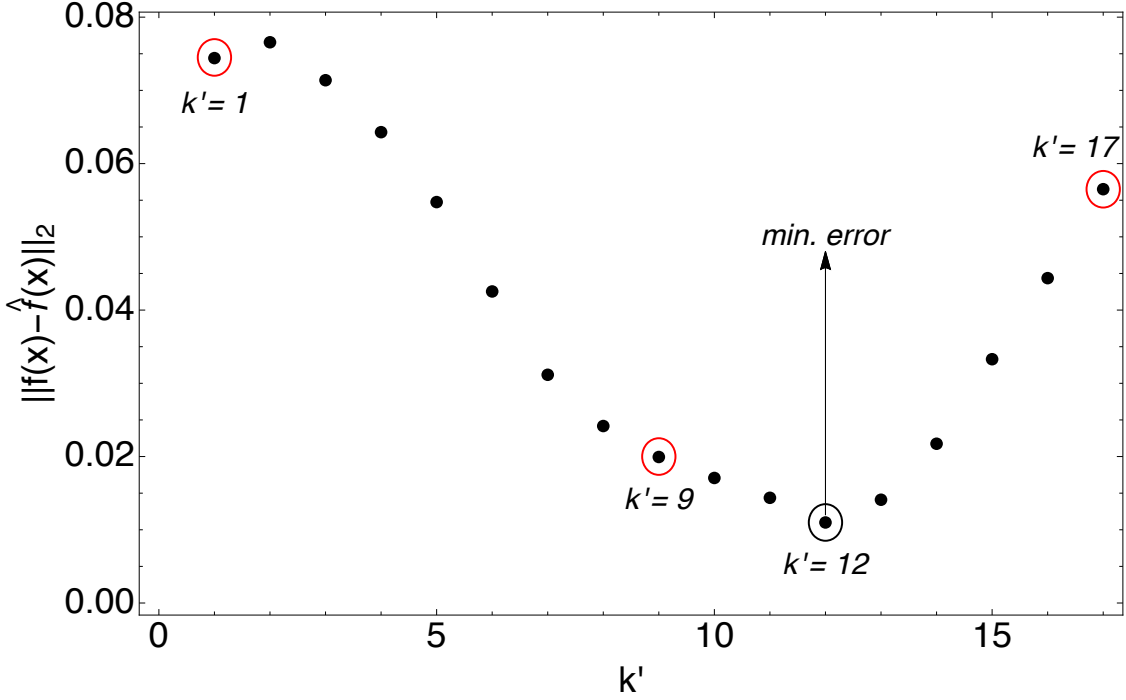} 
\caption{Plot of the \emph{L2 norm} as a function of $k^{\prime}$ such that the estimated $\widehat{f}(x)$ is calculated using $\beta_k$ for $k = k_{min} + k^{\prime}$. The points encircled in red are the 3 $k^{\prime}$s chosen for Figure~\ref{fig:rxerr}.}
\label{fig:ferr} 
\end{figure}

\begin{figure}[ht!]
\centering
\subfloat[Part 1][]{
\includegraphics[width=0.3\textwidth]{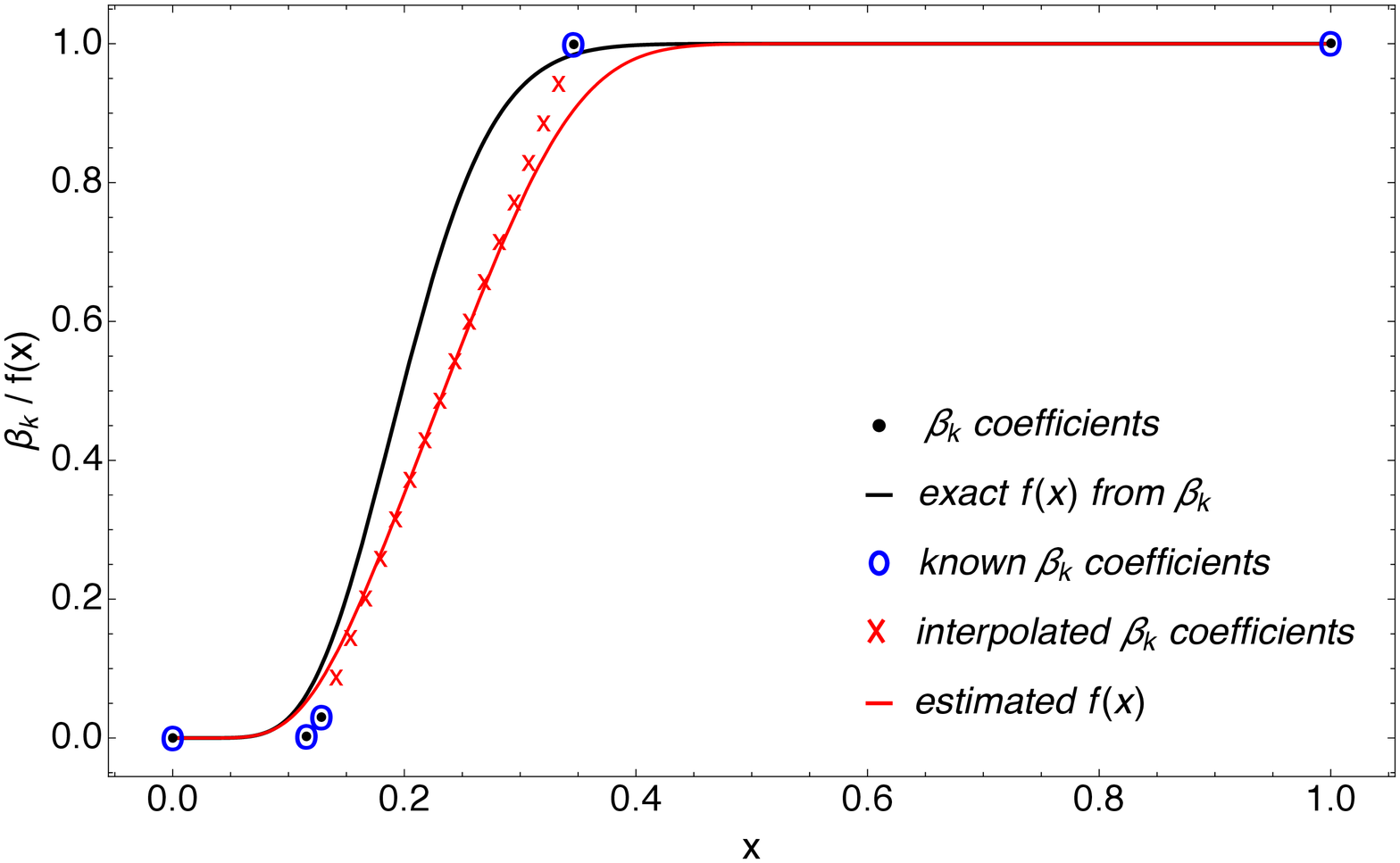} \label{fig:rxerr1} }
\subfloat[Part 1][]{
\includegraphics[width=0.3\textwidth]{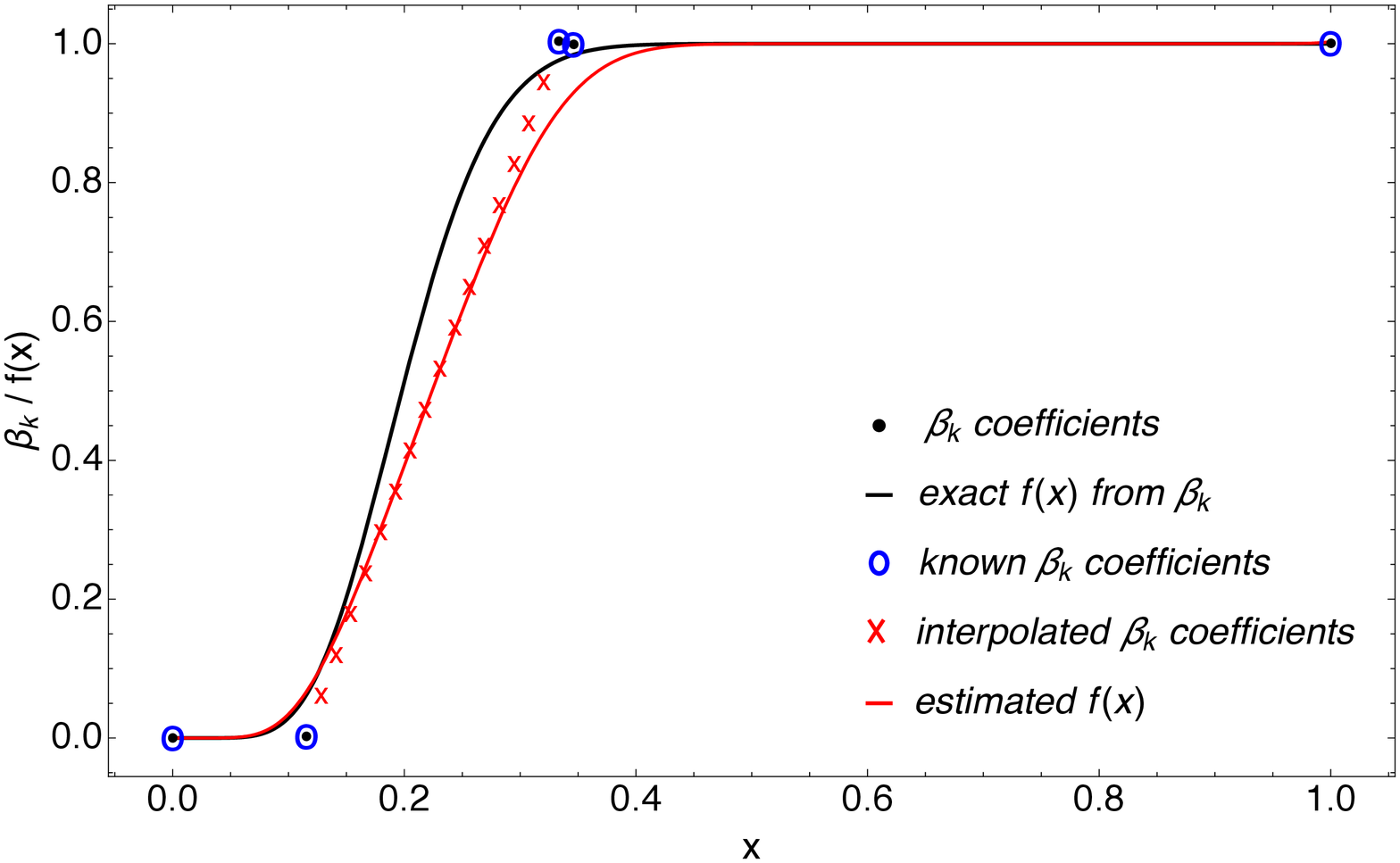} \label{fig:rxerr2} }
\subfloat[Part 1][]{
\includegraphics[width=0.3\textwidth]{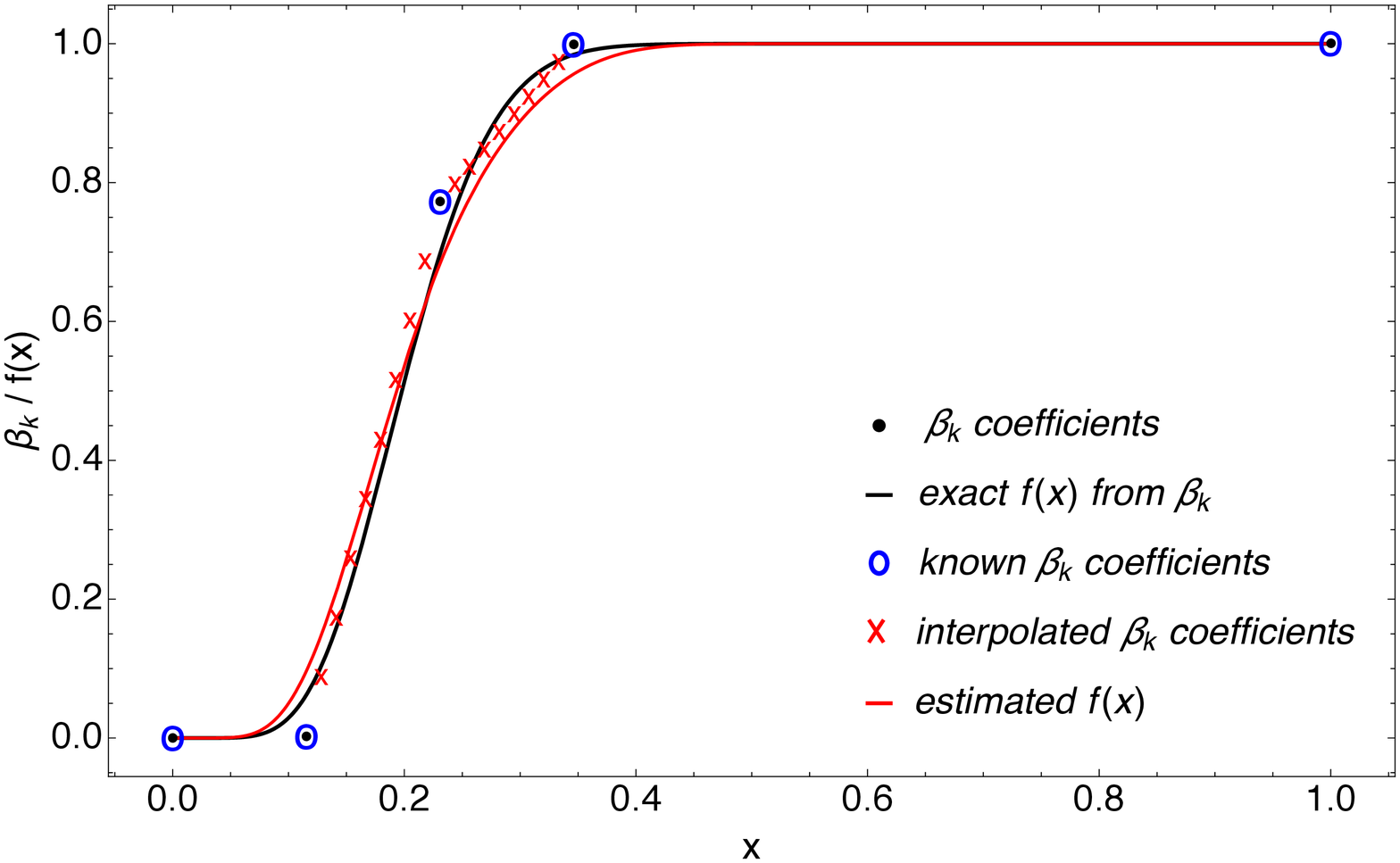} \label{fig:rxerr3} }
\caption{Estimated $\widehat{f}(x)$ using just 1 additional $(k, \beta_k)$. Results when (a) $k = k_{min} + 1$, (b) $k = k_{min} + 17 = k_{max} - 1$ and (c) $k = k_{min} + 8 = (k_{min}+k_{max})/2$.}
\label{fig:rxerr}
\end{figure}

Now the aim is to determine which measurement by the oracle will most reduce the error in the estimation. Figure~\ref{fig:ferr} shows the plot of the \emph{L2 norm} obtained for the different $\widehat{f}(x)_{k_{min}+k^{\prime}}$ as a function of $k^{\prime}$ for this case, where $k^{\prime} = 0,1, \ldots, n$.
For example, for the function $f(x)$ described in Section~\ref{sec:kar}, $N = 78$, $(k_{min}, \beta_{k_{min}}) = (9, 0.002)$ and $(k_{max}, \beta_{k_{max}}) = (27, 0.999)$ . The estimated $\widehat{f}(x)$ for three possible choices of measurement, $k = k_{min} + 1$, $k = k_{max} - 1 = k_{min} + 17$ and $k = (k_{min} + k_{max})/2 = k_{min} + 9$, are shown in Figures~\ref{fig:rxerr1}, \ref{fig:rxerr2} and \ref{fig:rxerr3}, respectively.

By def\/inition, we have the exact values $\beta_k = 0,  \forall \beta_k < \beta_{k_{min}}$ and  $\beta_k = 1, \forall \beta_k > \beta_{k_{max}}$. In practice, the ``oracle" is often a Monte Carlo simulation, and the cost $C$ of making a measurement can be exponential in the precision required. The cost may also depend on the value of $k$, but we ignore that possibility here. Assuming that errors in $\beta_k$ are independent and identically distributed, we can estimate the distribution of $\|f - \widehat{f} \|_2$ by Monte Carlo. This gives the likelihood, $\mathcal{L}$ that the observed data would be generated by the model $\widehat{f}(x)$. To calculate $ln (\mathcal{L})$, we draw 10 random samples of $\beta_k$ values lying in the range ($\beta_k - error$, $\beta_k+error$) from a binomial distribution (the sampling distribution of the Monte Carlo process used by the oracle) and obtain 10 different $f(x)$ curves. We use the following equation to determine $\mathcal{L}$.
\begin{equation}
-ln \mathcal{L} = \frac{1}{2} \frac{ \int_0^1 \left[ \sum_{k=0}^{N} \left( \widehat{\beta}_k - \widetilde{\beta}_k \right) B(N, k, x) \right]^2 dx}{\sigma^2}
\end{equation}
$\widehat{\beta}_k$'s are the estimated coef\/f\/icients obtained using linear interpolation, $\widetilde{\beta}_k$'s and $\sigma$ are the mean coef\/f\/icients and the standard deviation respectively obtained from the 10 samples.

We consider the addition of more $\beta_k$s as different models in this analysis and the one with the minimum $AIC$ and $BIC$ is chosen as the best candidate for f\/itting $f(x)$. Our models satisfy constraints on $f(x)$, the 4 basic parameters and some $m$ additional known $(k, \beta_k)$. Therefore, the number of parameters estimated for each model is given by $n - m$, where $m = 0, 1,2, \ldots, n$. The $AIC$ and $BIC$ values are calculated using
\begin{align}
AIC & =  2n - 2 ln (\mathcal{L}) \\
BIC & = ln(N)*n - 2 ln (\mathcal{L})
\end{align}
where $N$ is the total number of points and $\mathcal{L}$ is the maximum value of the likelihood function.

\subsection{Toy Example} \label{sec:toy}
Consider the non-decreasing monotonic degree-7 polynomial \footnote {This is the $ST$ reliability \citep{Moore:56} for the directed network $\lbrace S \rightarrow 1, 1 \rightarrow 2, 2 \rightarrow T, 1 \rightarrow 3, 3 \rightarrow T, S \rightarrow 4, 4 \rightarrow T \rbrace$, i.e., the probability that a message from node S will reach T if the probability of transmission is $x$ through each of the edges.}, 
\begin{equation}
f(x) = x^2 + 2 x^3 - 3 x^5 + x^7 
\label{eqn:toy}
\end{equation}
which satisf\/ies the boundary conditions, $f(0) = 0$ and $f(1) = 1$. We can express it in the Bernstein basis as
\begin{equation}
f(x) = \vec{\beta}_k \binom{7}{k} x^k (1-x)^{7-k}
\end{equation}
where $\vec{\beta}_k = \lbrace 0, 0, \frac{1}{21}, \frac{1}{5}, \frac{18}{35}, \frac{19}{21}, 1, 1 \rbrace$.

\begin{figure}[ht!]
\centering
\subfloat[Part 1][]{
\includegraphics[width=0.4\textwidth]{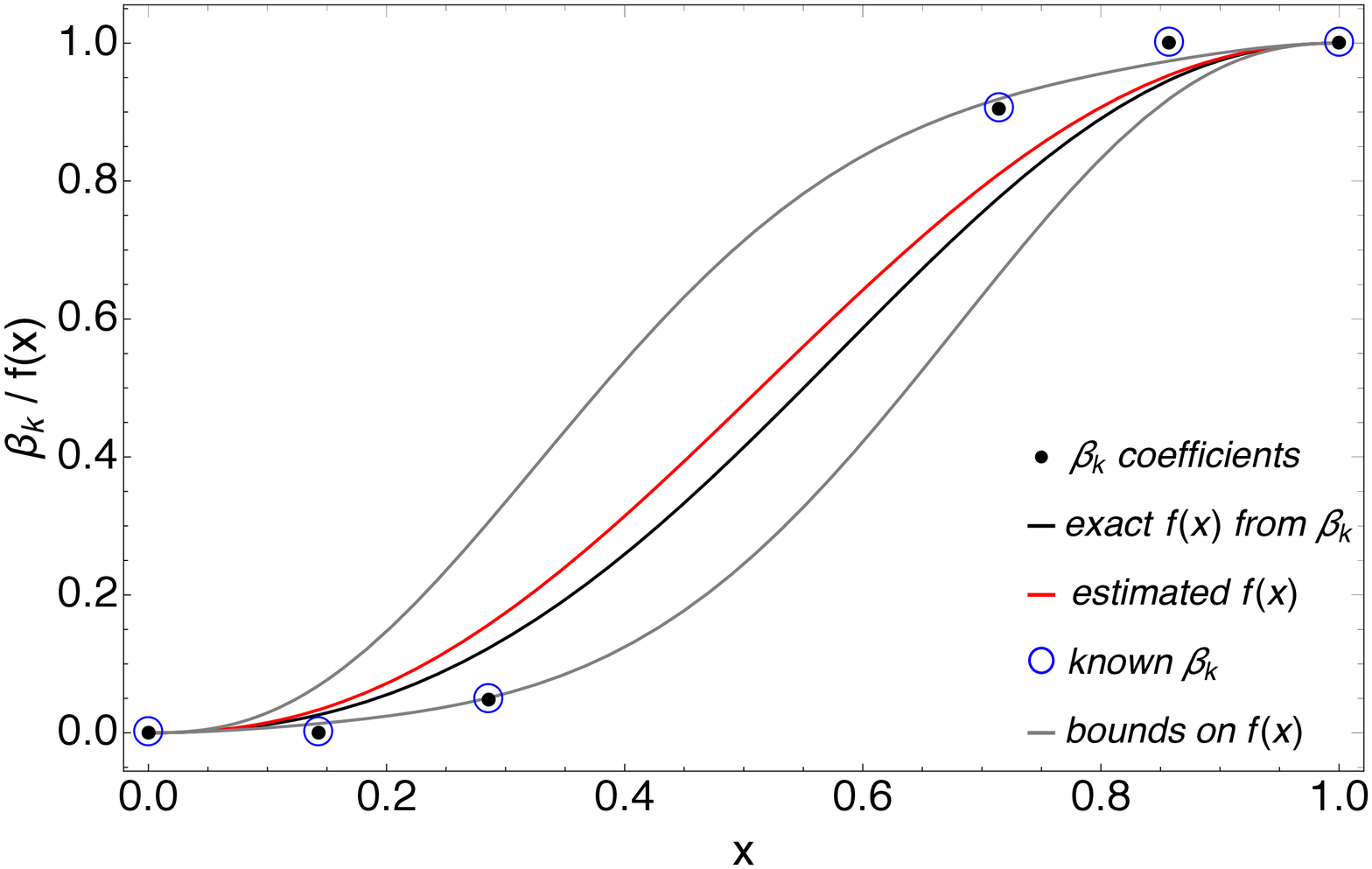} \label{fig:bnd1} }
\subfloat[Part 1][]{
\includegraphics[width=0.4\textwidth]{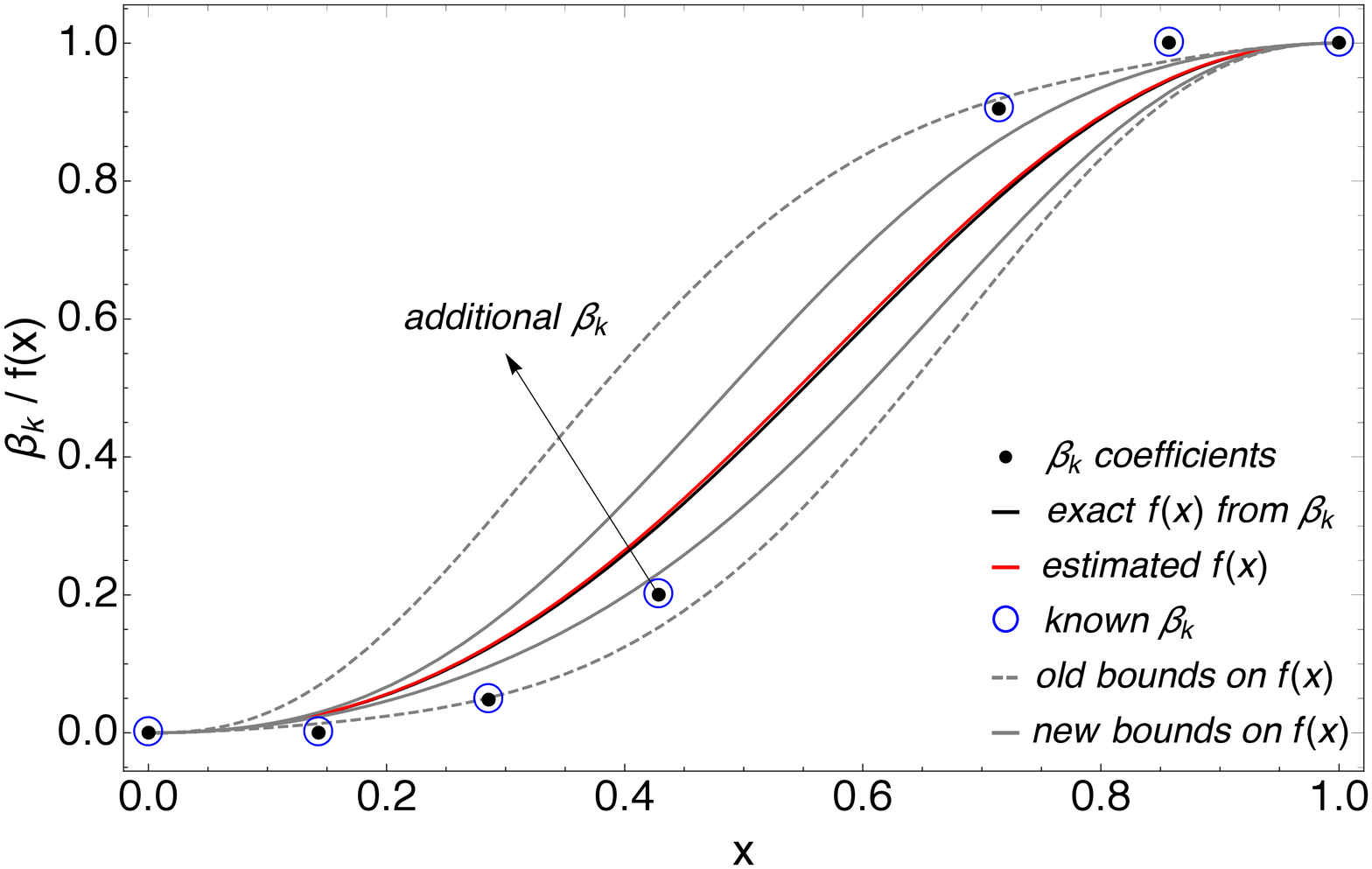} \label{fig:bnd2} } \\
\subfloat[Part 1][]{
\includegraphics[width=0.4\textwidth]{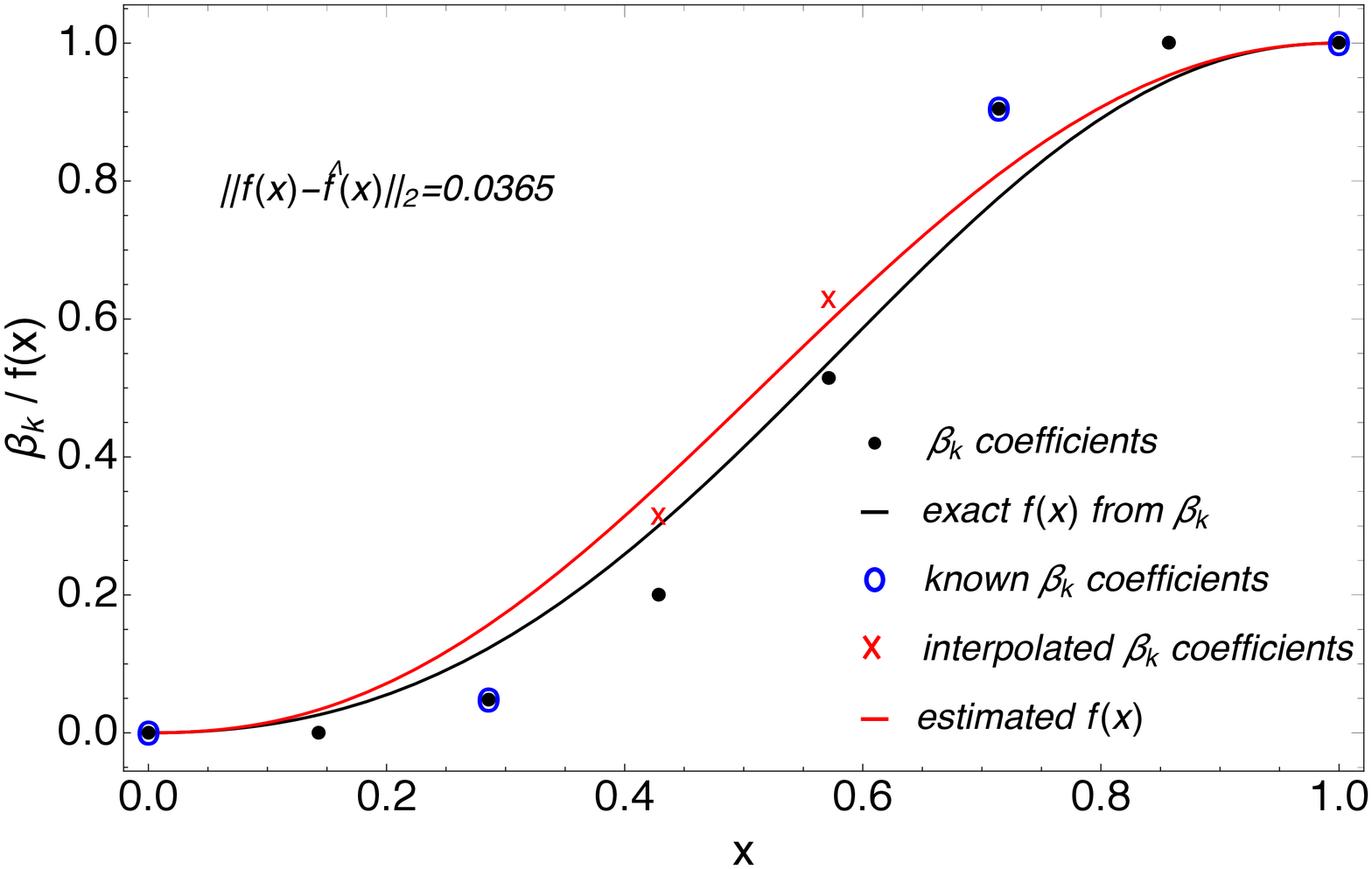} \label{fig:toy1} } 
\subfloat[Part 1][]{
\includegraphics[width=0.4\textwidth]{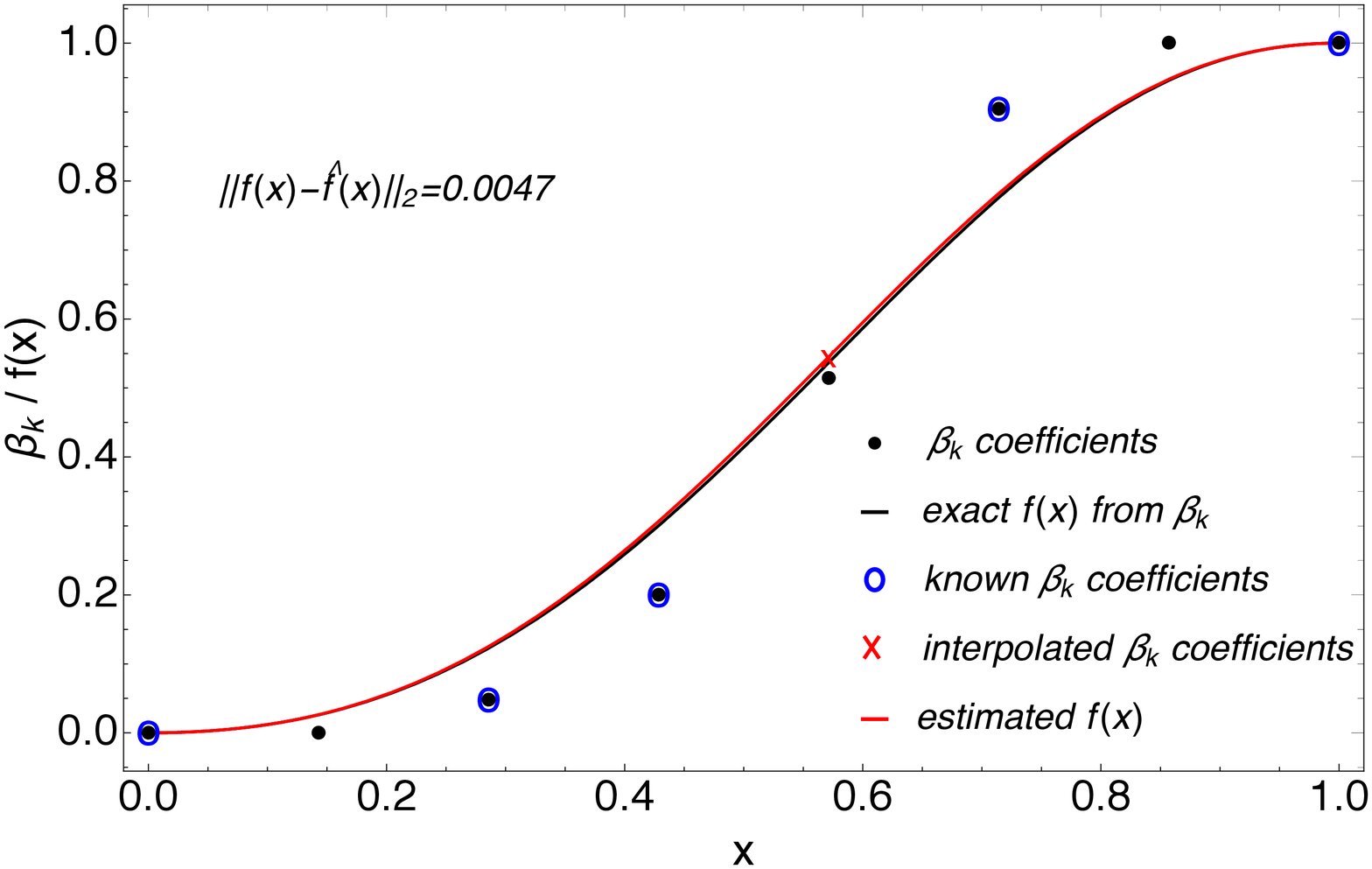} \label{fig:toy2} } \\
\subfloat[Part 1][]{
\includegraphics[width=0.4\textwidth]{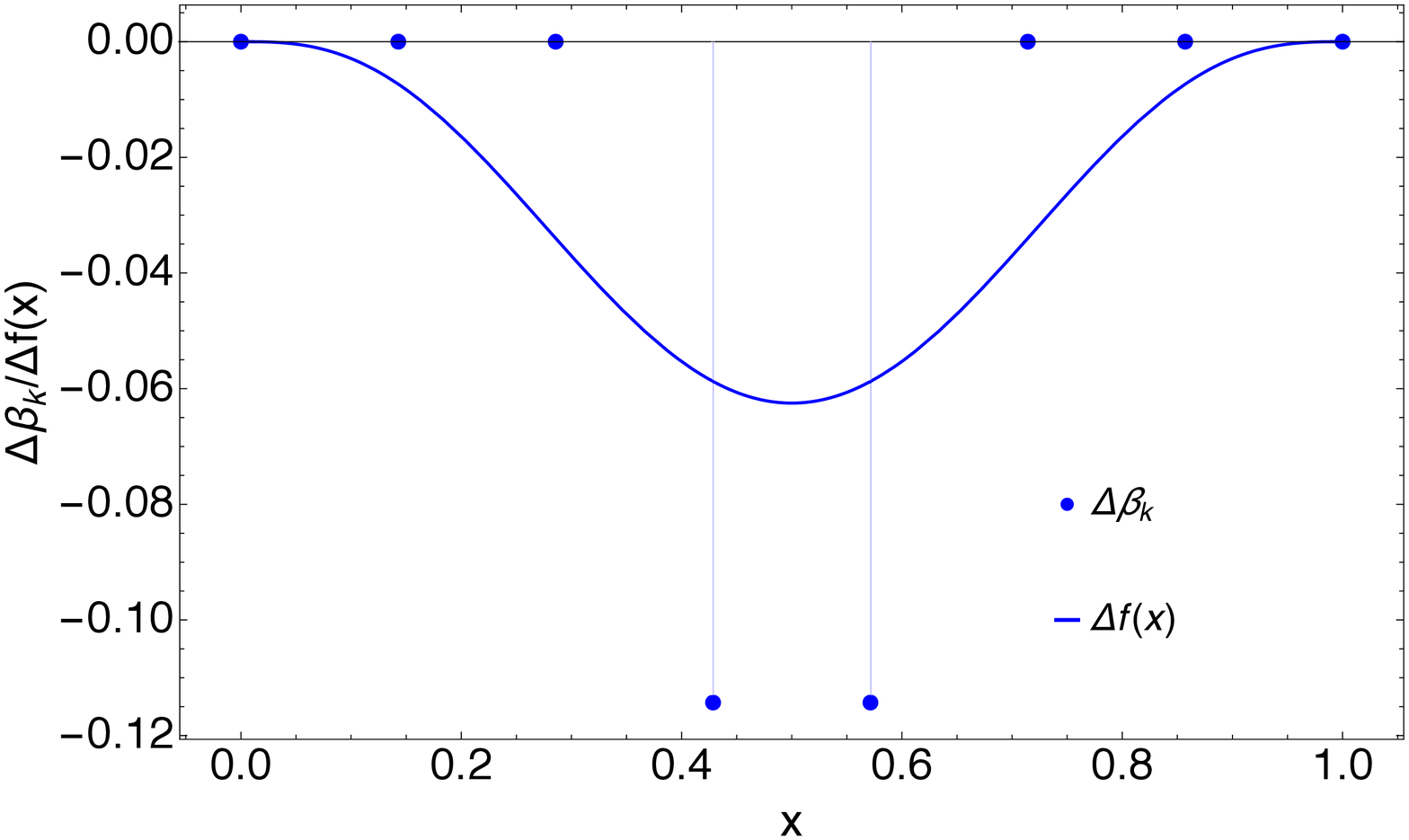} \label{fig:df1} }
\subfloat[Part 1][]{
\includegraphics[width=0.4\textwidth]{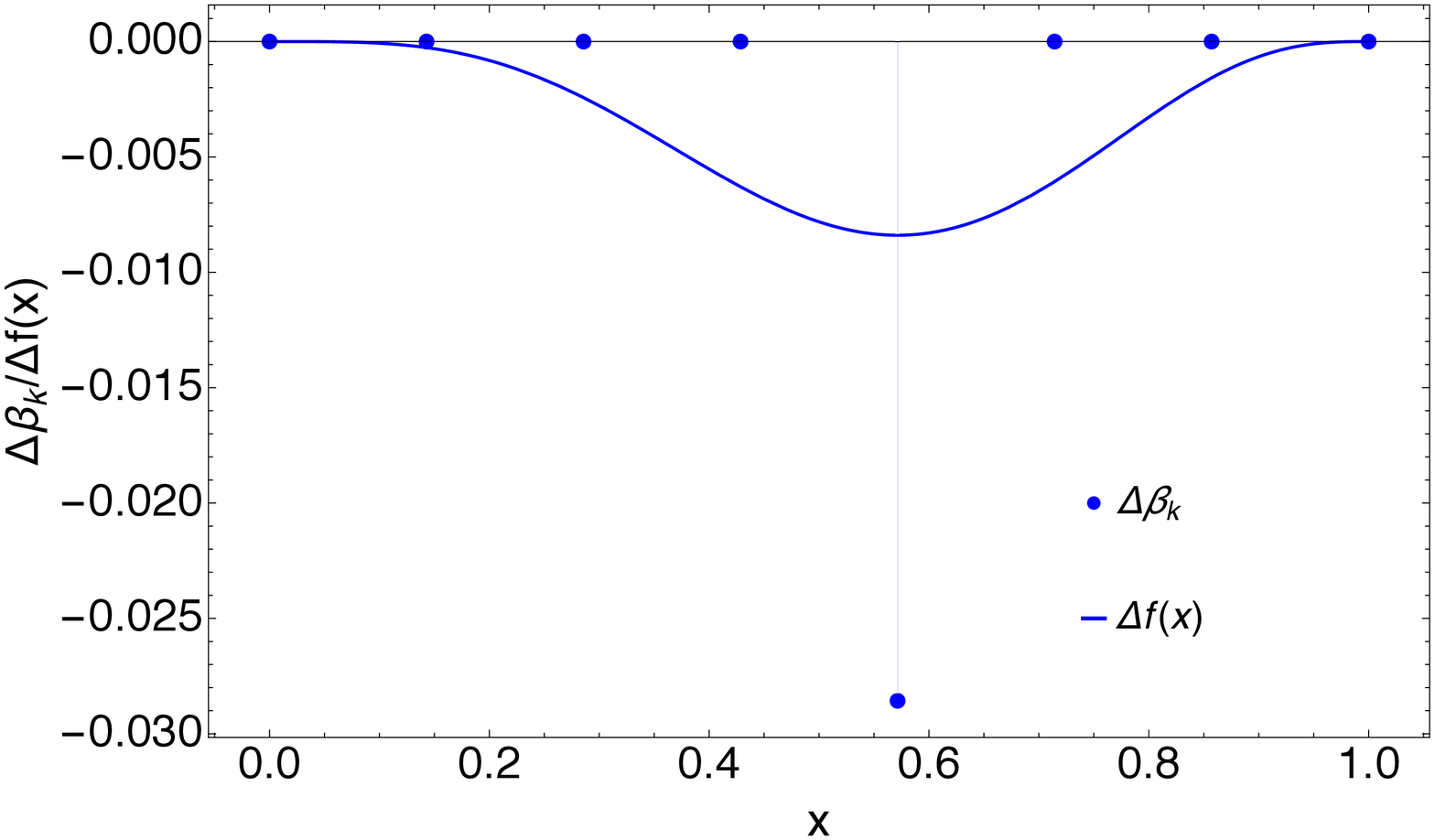} \label{fig:df2} }
\caption{Plot of $\beta_k$ and $f(x)$ as a function of $x$. The black solid line represents the exact $f(x)$. The red curve is $\widehat{f}(x)$. (a), (c) Bounds and $\widehat{f}(x)$ obtained knowing only $(k_{min}, \beta_{k_{min}})$ and $(k_{max}, \beta_{k_{max}})$. (b), (d) Bounds and $\widehat{f}(x)$ improve with the knowledge of one additional $\beta_k$. (e), (f) The difference between the exact and interpolated $\beta_k$ and $f(x)$. }
\label{fig:bnd}
\end{figure}

We start with the assumption that $(k_{min}, \beta_{k_{min}})$ and $(k_{max}, \beta_{k_{max}})$ are known. Figure~\ref{fig:bnd1} shows the bounds obtained from these known parameters. $f(x)$ and $\widehat{f}(x)$ must lie within these bounds. When we call the oracle to f\/ind one additional $\beta_k$, both the f\/it and the bounds improve as shown in Figure~\ref{fig:bnd2}. Figures~\ref{fig:toy1} and \ref{fig:toy2} show that $\widehat{f}(x)$ improves with the additional $\beta_k$ and $\|f(x) - \widehat{f}(x)\|_2$ decreases from 0.0365 to 0.0047. The red points in Figure~\ref{fig:toy1} and \ref{fig:toy2} represent the unknown linearly interpolated $\widehat{\beta}_k$s which yield the red $\widehat{f}(x)$. The difference in the estimated $\beta_k$'s and $f(x)$ with the exact ones are presented in Figure~\ref{fig:df1} and \ref{fig:df2}. 

\begin{figure}[ht!]
\centering
\includegraphics[width=\linewidth]{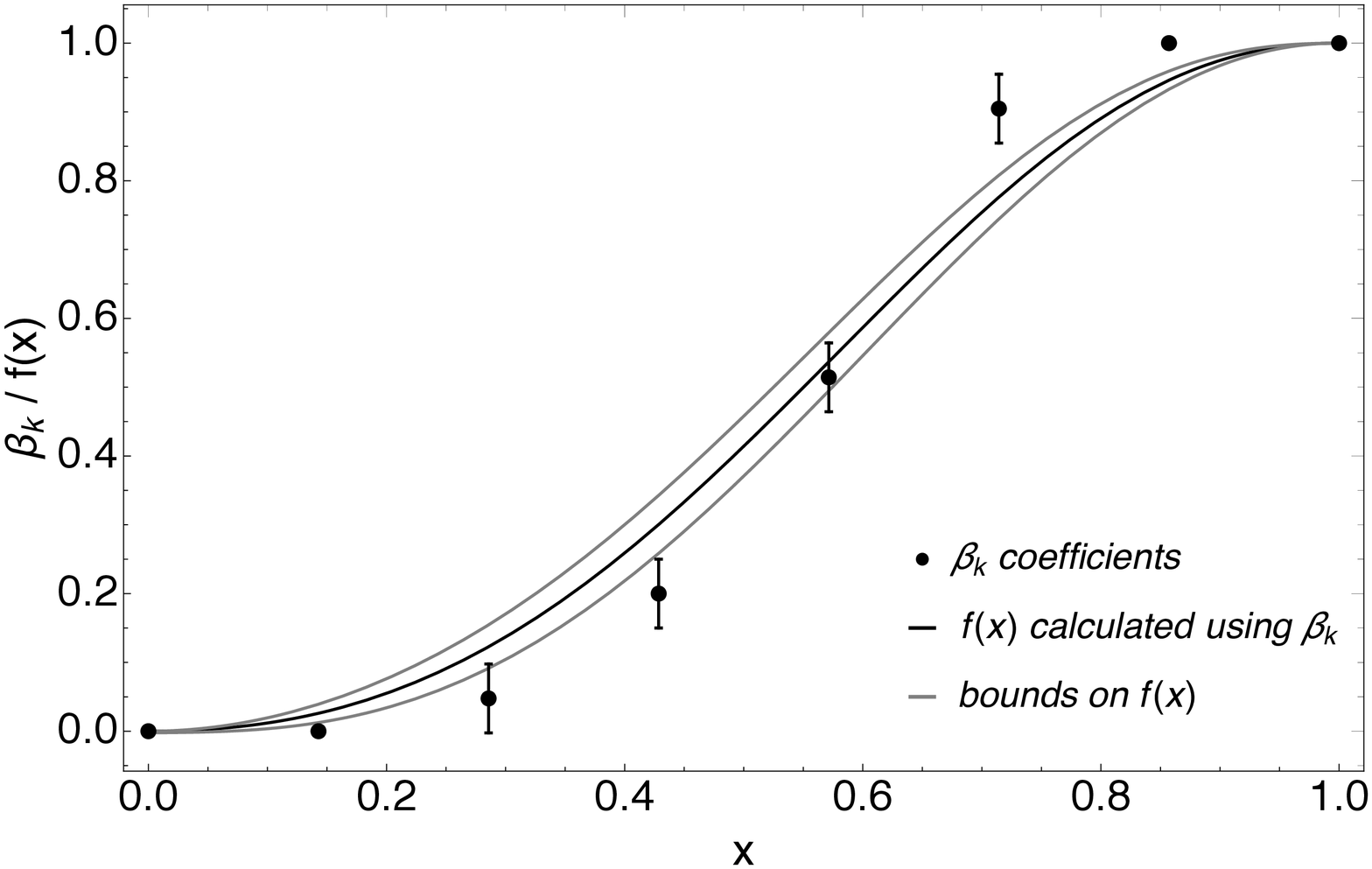} 
\caption{Bounds on $f(x)$, assuming uniform errors of up to 0.05 in the measurement of those $\beta_k$s which are not 0 or 1.}
\label{fig:bnd3}
\end{figure}

The $\beta_k$'s for this small network can be evaluated exactly. However, if there were an error associated with the estimation of the $\beta_k$s (those which are non-zero and not one), 
it could be handled by choosing random sets of $\beta_k$s from the sampling distribution, as above.
For example, if the errors are distributed uniformly in an interval around the observed value, the estimated $f(x)$ would lie between the bounds shown in Figure~\ref{fig:bnd3}. 

\subsection{Karate Network} \label{sec:kar}
\begin{figure}[ht!]
\centering
\includegraphics[width=\linewidth]{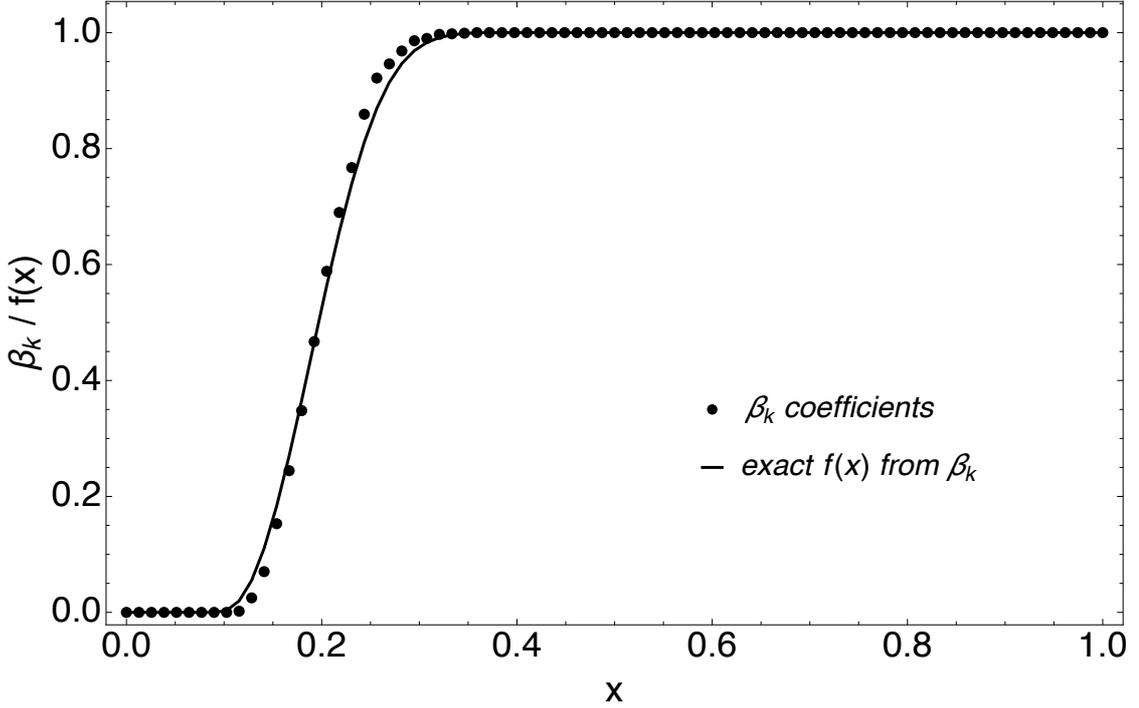} 
\caption{Plot of $\beta_k$ and $f(x)$ as a function of $x$ when connected sub-graphs contain 30\% of the vertices. The solid curve represents the $f(x)$ values calculated using all the $\beta_k$'s. The error in the estimates of $\beta_k$s is 0.005, which are of the size of the points.}
\label{fig:pkrx}
\end{figure}

The Zachary Karate club network \citep{zachary1977information} is a social network of a university karate club, which has 34 vertices and 78 edges. This network has been extensively studied in the literature. Here we consider $f(x)$ to be the probability that an infectious disease seeded in one randomly chosen person would eventually infect at least 30\% of this population (i.e., at least 10 others), when the person-person transmission probability is $x$ \footnote{This $f(x)$ is also a network reliability polynomial, which is related to the \emph{All Terminal} reliability  \citep{Moore:56, PhysRevE.88.052810, NATH2018121}.}. Figure~\ref{fig:pkrx} shows a high-precision Monte Carlo estimate of $f(x)$. For this network and this def\/inition of reliability, $k_{min} = 9$, $k_{max} = 27$ and $N = 78$. There are $n=17$ unknown $\beta_k$ lying between $\beta_{k_{min}}$ and $\beta_{k_{max}}$.

\begin{figure}[ht!]
\centering
\subfloat[Part 1][]{
\includegraphics[width=0.4\textwidth]{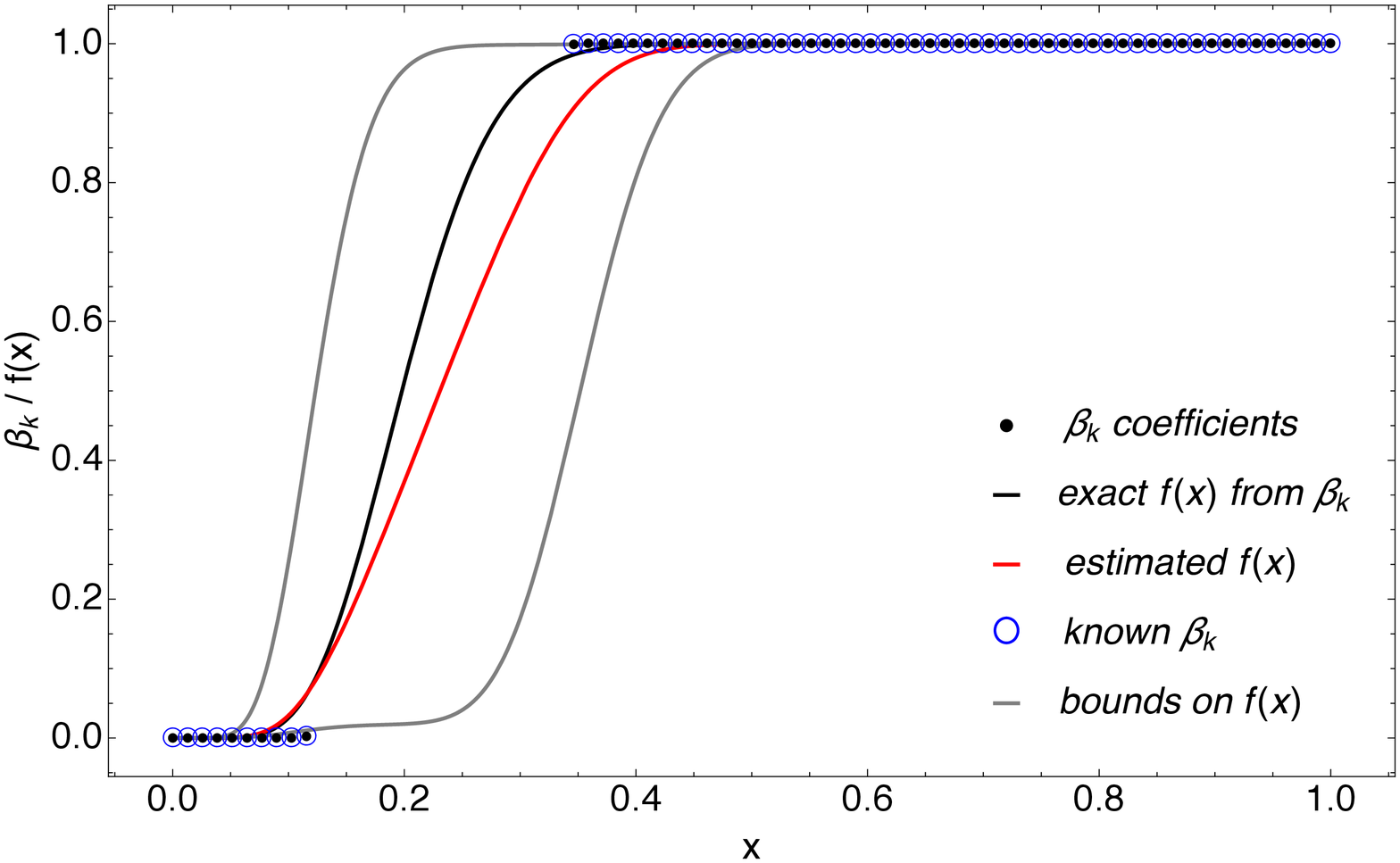} \label{fig:bk1} }
\subfloat[Part 1][]{
\includegraphics[width=0.4\textwidth]{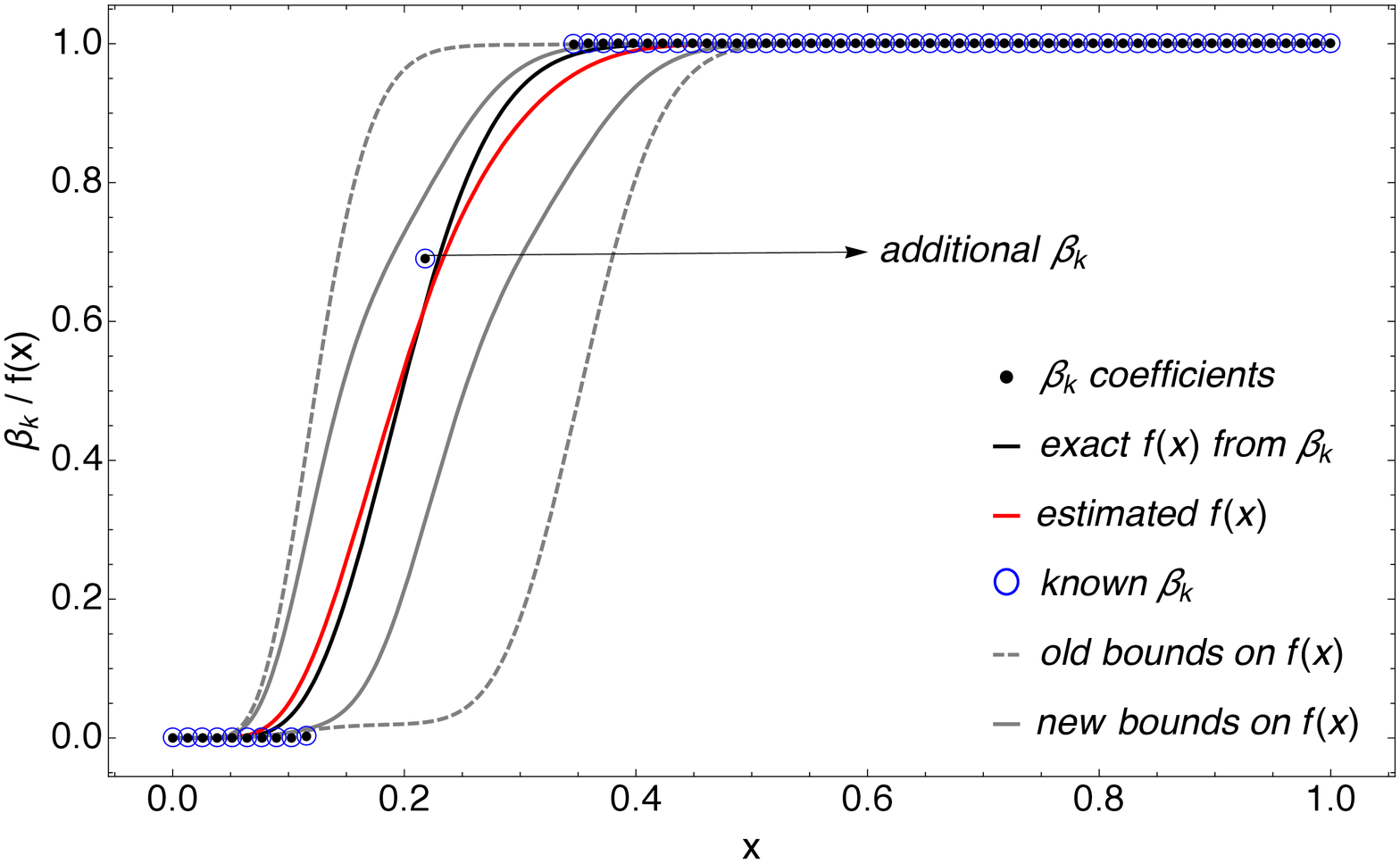} \label{fig:bk2} } \\
\subfloat[Part 1][]{
\includegraphics[width=0.4\textwidth]{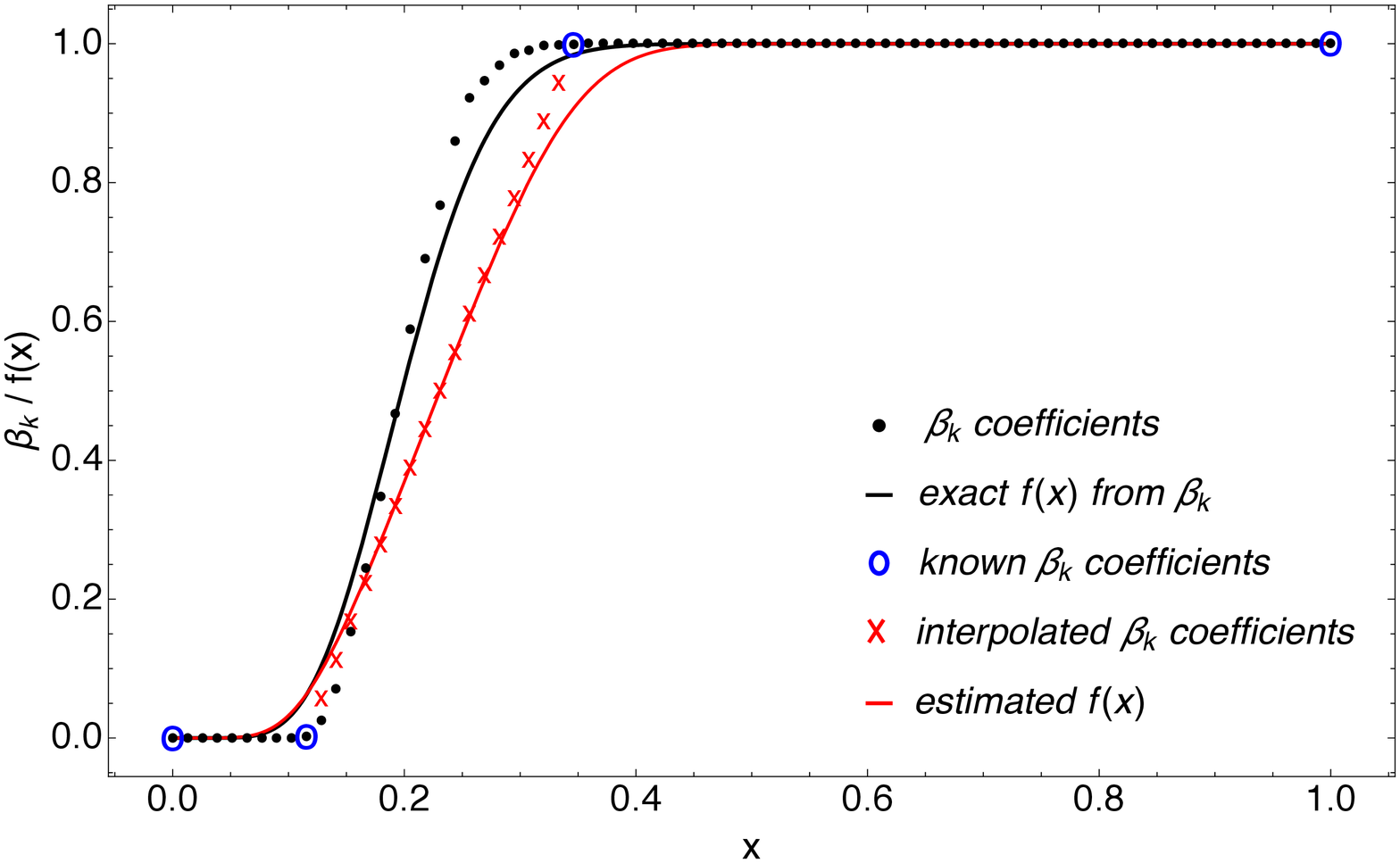} \label{fig:est} }
\subfloat[Part 1][]{
\includegraphics[width=0.4\textwidth]{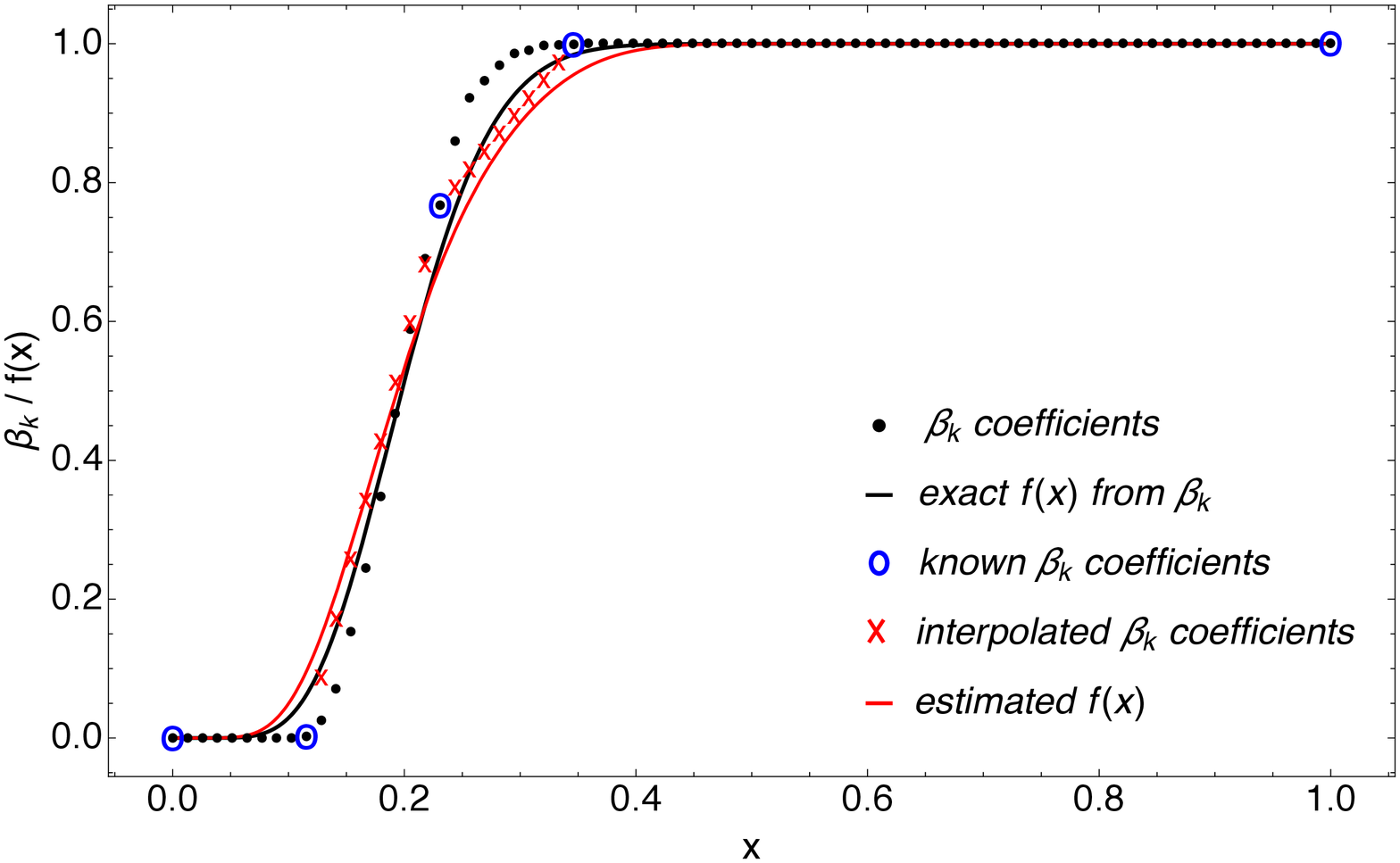} \label{fig:est1} } \\
\subfloat[Part 1][]{
\includegraphics[width=0.4\textwidth]{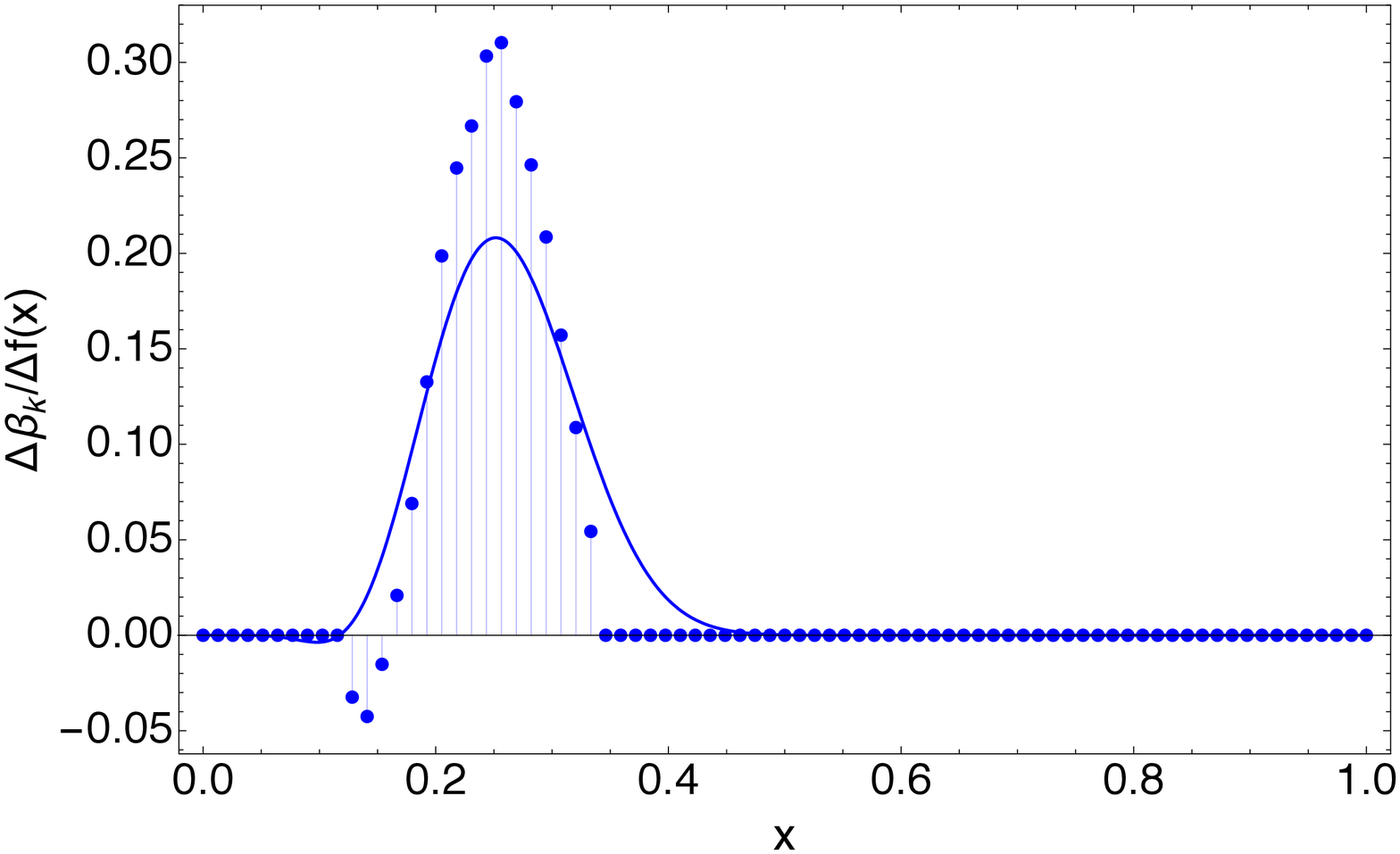} \label{fig:prdiff} }
\subfloat[Part 1][]{
\includegraphics[width=0.4\textwidth]{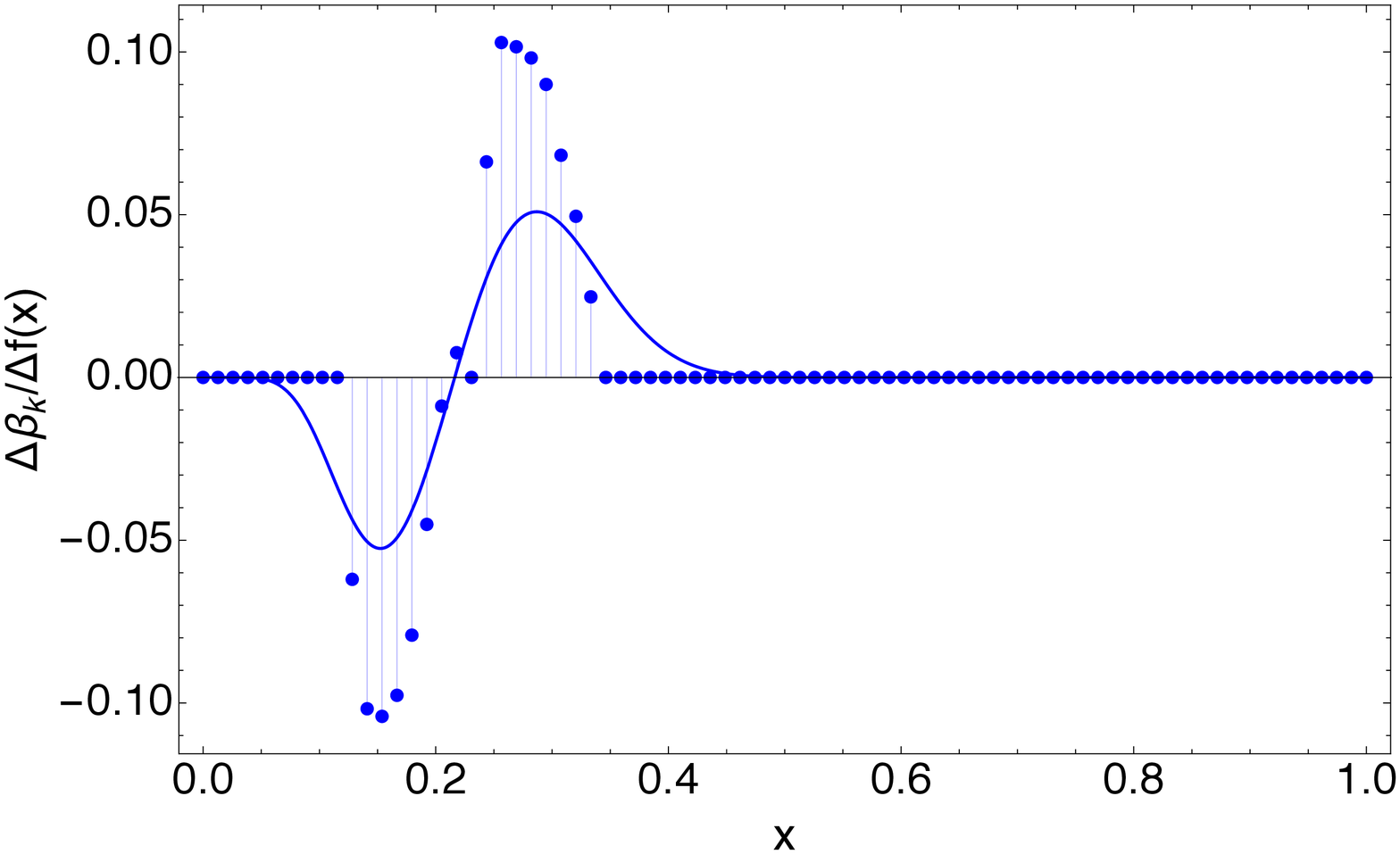} \label{fig:prdiff1} }
\caption{Plot of $\beta_k$ and $f(x)$ as a function of $x$. The black solid line represents the exact $f(x)$. The red curve is $\widehat{f}(x)$. (a), (c) Bounds and $\widehat{f}(x)$ obtained knowing only $(k_{min}, \beta_{k_{min}})$ and $(k_{max}, \beta_{k_{max}})$. (b), (d) Bounds and $\widehat{f}(x)$ improve with the knowledge of one additional $\beta_k$. (e), (f) The difference between the exact and interpolated $\beta_k$ and $f(x)$. As in Figure~\ref{fig:bnd}, one additional known $\beta_k$ signif\/icantly improves the bounds on $f(x)$. These results are obtained from model 1, i.e., $m = 0$ and $n = 17$.}
\label{fig:bern}
\end{figure}

Table~\ref{tab:model} summarizes the 6 models used for the analysis. The number of parameters estimated for each model, $n$, is given by $17-m$, where $m = 0, 1, 3, 5, 7, 9$ for this case. Similar to the example in Section~\ref{sec:toy}, Figure~\ref{fig:bern} shows that knowing one more $\beta_k$ along with the initial 4 parameters - $(k_{min}, \beta_{k_{min}})$ and $(k_{max}, \beta_{k_{max}})$, improves the estimate and the bounds of $\widehat{f}(x)$ and that it makes a difference which $\beta_k$ is measured. The results for models 2-6 are in the Supplementary Material. Table~\ref{tab:model} records the $\|f(x) - \widehat{f}(x)\|_2$, $AIC$ and $BIC$ values for these models.

\begin{table}
%% Caption MUST come immediately after \begin{table}
\caption{\label{tab:model}The L2 norm, AIC and BIC values for the 6 models used.}
\centering
\fbox{%
\begin{tabular}{*{7}{c}}
Model & Estimator & $m$ $(k, \beta_k)$ & $n$ & $\|f(x)-\widehat{f}(x)\|_2$ & $AIC$ & $BIC$ \\\hline
model 1 & $\widehat{f}_0$ & 0 & 17 & 0.0694 & $2.8505 \times 10^{8}$ & $2.8505 \times 10^{8}$ \\
model 2 & $\widehat{f}_1$ & 1 & 16 & 0.0201 & $2.3975 \times 10^{7}$ & $2.3975 \times 10^{7}$ \\
model 3 & $\widehat{f}_3$ & 3 & 14 & 0.0074 & $3.2294 \times 10^6$ & $3.2294 \times 10^6$ \\
model 4 & $\widehat{f}_5$ & 5 & 12 & 0.0066 & $2.6195 \times 10^6$ & $2.6196 \times 10^6$ \\
model 5 & $\widehat{f}_7$ & 7 & 10 & 0.0012 & $8.1958\times 10^4$ & $8.1982 \times 10^4$ \\
model 6 & $\widehat{f}_9$ & 9 & 8 & 0.0009 & $5.2715 \times 10^4$ & $5.2734 \times 10^4$ 
\end{tabular}
}
\end{table}

\section{Discussions}
Even when $f(x)$ is unknown, we can bound it with the given information. Figures~\ref{fig:ferr} and \ref{fig:rxerr} show that the choice of $k$ at which to evaluate $\beta_k$ affects the quality of $\widehat{f}(x)$. Figure~\ref{fig:ferr} also shows that if we could evaluate only one $\beta_k$ to estimate $f(x)$, we should pick $k = k_{min} + 12$ for this particular example because it reduces the \emph{L2 norm} the most. Figure~\ref{fig:bnd} shows that the bounds on $f(x)$ as well as the f\/it improve with additional measurements. The known $\beta_k$s are used to calculate the unknown ones by linear interpolation. The estimated $\widehat{f}(x)$ is obtained as a B\'{e}zier curve using these $\widehat{\beta}_k$s. The black solid curve is the exact $f(x)$ calculated using all the $\beta_k$ coef\/f\/icients and the red $\widehat{f}(x)$ is present within the bounds def\/ined by the initial 4 parameters - $(k_{min}, \beta_{k_{min}})$ and $(k_{max}, \beta_{k_{max}})$ and the additional $m$ $(k, \beta_k)$.  This is demonstrated for a system with larger $N$ in Figure~\ref{fig:bern}. 

There will be an estimation error associated with the $\widehat{\beta}_k$s obtained from Monte Carlo simulations for larger systems. Figure~\ref{fig:bnd3} show such an instance for the toy example when uniform error of 0.05 is assumed for all the $\beta_k$s which are neither 0 nor 1. The resultant $\widehat{f}(x)$ would lie within the bounded region. The error can be reduced by increasing the number of samples for the simulation yielding tighter bounds but this incurs possibly exponentially higher costs. Random samples for $\beta_k$s obtained from the binomial Monte Carlo sampling distribution are used to calculate the likelihood that the true data has been generated by $\widehat{f}(x)$. The difference in the $\beta_k$ values from the random trials is plotted in the Supplementary Material.

The exact and estimated $f(x)$ and the $\beta_k$'s for the Karate network for the various models listed in Table~\ref{tab:model} are plotted in Figure~\ref{fig:bern}. The $\|f(x) - \widehat{f}(x)\|_2$ in Table~\ref{tab:model} decreases with the decrease in the number of estimated parameters. Since the standard deviation from the data is so small, the $ln (\mathcal{L})$ term dominates the information criteria, the number of parameters is actually irrelevant and it doesn't matter which one of $AIC$ or $BIC$ is used. However, the change in $AIC$ or $BIC$ decreases monotonically with $m$. Even though we do not obtain a minimum, it is still useful to calculate them because we place a threshold on what kind of error is tolerated. In practice, for some $m$, it will drop below the threshold of $C$ bits per data point imposed by the cost of calling the oracle. For example, measuring only 7 of the 17 unknown $\beta_k$s in the Karate network gives a good estimate of its reliability, as long as the right 7 are measured.

\section{Conclusions}
Many questions about f\/inite, discrete stochastic systems can be answered by estimating the Moore Shannon network reliability, a probability distribution that is a high degree polynomial. Bernstein kernel density estimators provide extremely useful estimates for the overall reliability, when individual terms in the polynomial are estimated to arbitrary precision with Monte Carlo simulations. In addition, boundary conditions can often be used to specify many coef\/f\/icients exactly. The corresponding B\'{e}zier curves can incorporate both the boundary constraints and simulation results. These Bernstein polynomials provide tight, provable bounds as well as achieving good estimates. We have shown here how these bounds can be used to guide a practical adaptive measurement process that ef\/f\/iciently uses calls to the Monte Carlo oracle to create a uniformly good approximation. 

The number of samples required to reduce the error associated with precise measurements of the Bernstein coef\/f\/icients using Monte Carlo simulations increases exponentially with the system size. The interplay between increasing the precision in estimates of individual $\beta_k$s and increasing the number of $k$s at which the $\beta_k$s are measured remains to be investigated.

\section*{Acknowledgement}
The authors would like to acknowledge Dr. Patrick Huber, Department of Physics, Virginia Tech. We thank our external collaborators and members of the Network Dynamics and Simulation Science Laboratory (NDSSL) for their suggestions and comments. This work has been partially supported by Defense Threat Reduction Agency Comprehensive National Incident Management System Contract HDTRA1-17-0118, by the National Institute of General Medical Sciences of the National Institutes of Health under a Models of Infectious Disease Agent Study (MIDAS) Grant U01GM070694 and by the United States Agency for International Development Grant AID-OAA-L-15-0001. The content is solely the responsibility of the authors and does not necessarily represent the of\/f\/icial views of the National Institutes of Health, the Department of Defense or the United States Agency for International Development.

%Add funding details.

\bibliographystyle{rss}
\bibliography{draft}
\end{document}